\input harvmac
\input epsf

\def\w{\omega}
\def\h{{1\over2}}
\def\es{e^{2Q t}}
\def\et{e^{Q t}}
\def\p{\partial_{t}}
\def\em6{e^{-2Q t}}
\def\yiz{{Y_{0}}^i}
\def\yia{{Y_{a}}^i}
\def\ynz{{Y_{0}}^9}
\def\yna{{Y_{a}}^9}
\def\yio{{Y_{1}}^i}
\def\yit{{Y_{2}}^i}
\def\yih{{Y_{3}}^i}
\def\yno{{Y_{1}}^9}
\def\ynt{{Y_{2}}^9}
\def\ynh{{Y_{3}}^9}
\def\yoo{{Y_{1}}^1}
\def\yot{{Y_{2}}^1}

\def\ao{A_{1}}
\def\at{A_{2}}
\def\ah{A_{3}}
\def\tao{{A^+}_{1}}
\def\tat{{A^+}_{2}}
\def\tyo{{A^-}_{1}}
\def\tyt{{A^-}_{2}}
\def\inzer{{\int_{0}}^\infty}
\def\ininf{{\int^\infty_{-\infty}}}
\def\kab{K_{\alpha\beta}(t',t;s)}
\def\pf{{\pi\over4}}
\def\pic{({\pi c'\over2})}
\Title{}{\vbox{\centerline{A One Loop Problem of the Matrix Big
Bang Model}}} \centerline{Miao Li$^{1,2,3}$ and Wei Song$^{1,3}$}
\medskip
\centerline{\it $^1$ Institute of Theoretical Physics}
\centerline{\it Academia Sinica, P. O. Box 2735} \centerline{\it
Beijing 100080} \centerline{\it and}
\medskip
\centerline{\it $^2$ Interdisciplinary Center of Theoretical
Studies} \centerline{\it Academia Sinica, Beijing 100080, China}
\centerline{\it and}
\medskip
\centerline{\it $^3$ Interdisciplinary Center for Theoretical
Study} \centerline{\it University of Science and Technology of
China, Hefei, Anhui 230026, China}
\medskip

\centerline{\tt mli@itp.ac.cn} \centerline{\tt wsong@itp.ac.cn}
\medskip

We compute the one-loop effective action of two D0-branes in the
matrix model for a cosmological background, and find vanishing
static potential. However, there is a non-vanishing $v^2$ term not
predicted in a supergravity calculation. This term is complex and
signals an instability of the two D0-brane system, it may also
indicate that the matrix model is incorrect.

\Date{Dec. 2005}

\nref\csv{B. Craps, S. Sethi and E. Verlinde,``A Matrix Big Bang,"
hep-th/0506180.} \nref\mli{M. Li, ``A class of cosmological matrix
models," hep-th/0506260, Phys.Lett. B626 (2005) 202.}
 \nref\bchen{B.
Chen, ``The Time-dependent Supersymmetric Configurations in M-theory
and Matrix Models," hep-th/0508191, B. Chen.} \nref\hezh{ Y-l He and
P. Zhang, ``Exactly Solvable Model of Superstring in Plane-wave
Background with Linear Null Dilaton," hep-th/0509113.} \nref\ohta{T.
Ishino, H. Kodama and N. Ohta, ``Time-dependent Solutions with Null
Killing Spinor in M-theory and Superstrings'', hep-th/0509173,Phys.
Lett. B 631 (2005) 68.}
 \nref\sumit{S. R. Das and J. Michelson, ``pp Wave
Big Bangs: Matrix Strings and Shrinking Fuzzy Spheres,"
hep-th/0508068, Phys.Rev. D72 (2005) 086005.} \nref\berkooz{M.
Berkooz, Z. Komargodski, D. Reichmann and V. Shpitalnik, ``Flow of
Geometries and Instantons on the Null Orbifold," hep-th/0507067.}
\nref\she{ J.-H. She, ``A Matrix Model for Misner Universe and
Closed String Tachyons," hep-th/0509067 ;``Winding String
Condensation and Noncommutative Deformation of Spacelike
Singularity," hep-th/0512299.}  \nref\hikida{ Yasuaki Hikida, Rashmi
R. Nayak, Kamal L. Panigrahi, ``D-branes in a Big Bang/Big Crunch
Universe: Misner space," hep-th/0508003,JHEP 0509 (2005) 023.}
\nref\tai{  Yasuaki Hikida and Ta-Sheng Tai,``D-instantons and
Closed String Tachyon in Misner Space," hep-th/0510129.}
\nref\mls{M. Li, W. Song,``Shock waves and cosmological matrix
models," hep-th/0507185, JHEP 0510 (2005) 073.}\nref\bfss{T. Banks,
W. Fischler, S. Shenker and L. Susskind, ¡°M theory as A Matrix
Model: A Conjecture,¡± hep-th/9610043, Phys. Rev. D55 (1997) 5112.}
 \nref\abbott{L. F.
Abbott, ``The Background Field Theory Method Beyond One Loop," {\it
Nucl. Phys.} {\bf B185}(1981) 189.} \nref\becker{K.Becker and
M.Becker,``A Two-Loop Test of M(atrix) Theory",hep-th/9705091. }
\nref\gsrz{I.S.Gradshteyn and L.M.Ryzhik, ``Table of integrals,
series, and products".} \nref\stefan{S.~Fredenhagen, and
V.~Schomerus, ``On Minisuperspace Models of S-branes",
hep-th/0308205, JHEP 0312 (2003) 003. }
\newsec{Introduction}

Formulating string/M theory in a time-dependent background remains
an elusive problem. The only observables in string theory are
S-matrix elements, this is certainly true in the perturbative
formulation of string theory where conformal symmetry on the
world-sheet plays the role of the guiding principle in
constructing consistent asymptotically flat background, it is also
true in a nonperturbative formulation of M theory, the matrix
theory, where scattering amplitudes among D0-branes and their
bound states are assumed to exist. However, S-matrix does not
exist for most of interesting cosmological backgrounds, it
certainly does not exist for our universe. Perhaps, a
reformulation of observables is the key to extending string/M
theory to include time-dependent backgrounds.

The matrix model proposed by Craps et al. is an attempt to formulate
string theory in a time-dependent background \csv, the metric in
this model depends on a null coordinate and in the Einstein frame it
exhibits a null singularity at the ``big bang" point. This model was
subsequently generalized to a class of more general backgrounds in
\mli, and to a class of even more general backgrounds in \bchen,
\hezh, and \ohta\ (a concrete model in this class was previously
studied in detail in \sumit). For related work on time-dependent
backgrounds, see \berkooz, \she, \hikida, and \tai.

So far, except for the decoupling argument presented in \csv,
there has been no independent check on the correctness of the
matrix proposal. The effective action of a D0-brane in the
background generated by another D0-brane was derived in \mls,
where it is noticed that the usual double expansion in the
relative velocity $v$ and the inverse of the relative separation
$b$ fails when time is sufficiently close to the big bang point.
Although there is no definition of scattering amplitude between
two D0-branes too, we believed that it makes sense to talk about
the effective action at later times. In the present work we shall
make the usual one loop calculation to see whether we can obtain
the small velocity expansion of \mls. To our surprise, we shall
see that the $v^2$ term in the one-loop calculation does not
vanish and is complex. This is a rather astonishing result.

We are faced with two possibilities, our result may indicate that
the matrix proposal is incorrect, or it may signal an instability
of the two D0-brane system at later times, since the $v^2$ term in
the effective action is complex. However, we can not locate a
physical reason for this instability at present.

The layout of this paper is as follows. We use the background field
method of \abbott\ to write down a gauge-fixed action and expand it
to the second order. We compute the one-loop contribution of the
off-diagonal fluctuations to the effective action of two D0-branes
in sect.3 when the relative velocity vanishes, and find it equal to
zero. The $v^2$ term in the one-loop contribution is calculated in
sect.4, and we find a non-vanishing complex term. We show that the
small velocity expansion makes sense in the flat matrix theory and
the $v^2$ terms cancel in appendix A. Appendixes B is devoted to
discussions on the propagators.

\newsec{Basic setup}

In \csv, the authors consider a flat type IIA background, with a
null linear dilaton, $\phi=-Q x^+$. The Einstein metric has a
curvature singularity at $x^+=-\infty$. A matrix string action is
proposed in \csv\ to describe the theory nonperturbatively. The type
IIA background can be obtained by compactifying M theory on a
circle, along the ninth direction. In \mli, the background is lifted
to M theory, and the corresponding matrix theory is BFSS like \bfss.
D0 brane interaction is found out by considering the shock wave
solution in \mls. To get the shock wave solution, the authors have
compactified the ninth direction and averaged the source over that
direction. The Routhian of a graviton in the presence of another is
$\h p_- \sum_{n=1}^\infty
c_nv^2[\kappa_{11}^2e^{-2Qx^+}p'_-v^2r^{-6}]^{n-1}$, where $p_-$ is
the null momentum of the test particle, $(2\pi)^2 RR'p'_-$ is that
of the source particle, and $c_n$ is some fixed numerical
coefficient, especially, $c_1=1,c_2={1\over8\pi^2}$. $R$ is the
radius of the M-theory and $R'$ is that of 9th direction. From the
form of the Routhian, one can see that there is no static potential
between two gravitons, and there is no $v^2$ correction, either. In
this approach, we are expanding the effective action in terms of
$\kappa_{11}^2e^{-2Qx^+}p'_-v^2r^{-6}$. Note that
$\kappa^2={\kappa_{11}^2e^{-2Qx^+}\over2\pi
R'}={\kappa_{11}^2g_s^2\over 2\pi R'}$ is just the physical
gravitational constant in IIA string theory. Therefore it is clear
that the expansion is a supergravity perturbation.

In the present paper, we will just consider the case of two D0
branes, and hence $p_-={1\over R}$, and $p'_-={1\over R2\pi R 2\pi
R'} $. We shall in this paper work in the scheme of \mli, and use
the matrix
 model action
 instead of the matrix string action. We
compute the effective potential of two D0 branes with separation
both in the ninth direction and in the transverse directions.
 To compare our matrix model calculation with the supergravity
 result in \mls, the separation in the ninth direction should be
 integrated out in the end.

The matrix theory action includes the bosonic part $S_{B}$ and
fermionic part $S_{F}$. Set the Planck scale $l_{p}$ to $1$, the two
parts can be written as \eqn\sb{\eqalign{S_{B}&=\int d t
Tr\{{1\over2R}{(D_t X^i)}^2 +{1\over2R}e^{-2Qt}{(D_{\tau}X^9)}^2
+{R\over 4}e^{2Qt}{[X^{i},X^{j}]}^2 +{R\over 2}{[X^9,X^j]}^2\},\cr
S_{F}&=\int dt Tr \{i{\theta}^T D_t\theta -Re^{Qt}{\theta }^T
\gamma_{i}[X^i,\theta] -{R\theta}^T\gamma_{9}[X^9,\theta]\},}} where
$i,j=1,\cdots 8$, runs over the eight transverse directions, and
$D_t=\p+i[A,$ is the covariant derivative. Rescale
$t\rightarrow(2^{1/3}R)^{-1}t,Q\rightarrow2^{1/3}RQ$ and
$X^\mu\rightarrow 2^{1/3}X^\mu$ to absorb the $R$ in the action, we
have \eqn\sb{\eqalign{S_{B}&=\int dt Tr\{{(D_{t}X^i)}^2
 +e^{-2Q t}{(D_{t}X^9)}^2+{1\over 2}e^{2Q t}{[X^{i},X^{j}]}^2
+{[X^9,X^j]}^2\},\cr S_{F}&=\int dt Tr\{i\theta^T D_{t}\theta-e^{Q
t}\theta^T\gamma_{i}[X^i,\theta]-\theta^T\gamma_{9}[X^9,\theta]\}.}}
To calculate the effective potential, we use the background field
method \abbott. Expand the action \sb\ around the classical
background field $B^\mu$ by setting $X^\mu=B^\mu+Y^\mu$,
$\mu=1,2\cdots 9$. The fluctuation part of the action is a sum of
five terms \eqn\action{S=S_{i}+S_{9}+S_{A}+S_{fermi}+S_{ghost}.} In
the following, we will determine the explicit form of each term. It
is convenient to choose the gauge \eqn\guage{G\equiv\partial_{t}A-
ie^{2Q t} [B^i,X^i]-i[B^9,X^9]=0.} In the standard gauge fixing
procedure, we need to insert \eqn\unit{1=\Delta_{fp} \int [d\xi]
\delta(G-f(t)g(t))} into the path integral, where $\xi$ is a gauge
parameter, $f(t)$ is chosen to be $f(t)=e^{Q t}$ for later
convenience, $g(t)$ is any function. The path integral is
independent of the choice of $g(t)$, so we can multiply the path
integral by $\int [dg(t)]e^{-i {g(t)}^2}$. $\Delta_{fp}$ is given by
the variation of $G$ under gauge transformation, independent of
$g(t)$. Thus by changing the order of integration, we can integrate
out $g(t)$,
 and get a gauge fixing term
\eqn\sg{S_{gf}=- e^{-2Q t}G^2.} So the bosonic action of the
fluctuation is \eqn\si{\eqalign{S_{Y^i}=&  \int dt Tr
\{{(\partial_{t}Y^i)}^2 +\es
({[B^i,Y^j]}^2+\em6[B^9,Y^j]^2+{[B^i,Y^i]}^2\cr
&+2[B^i,Y^j][Y^i,Y^j]+\h {[Y^i,Y^j]}^2)\},\cr S_{Y^9}=&  \int dt\em6
Tr\{ {(\partial_{t}Y^9)}^2 +\es({[B^i,Y^9]}^2+{\em6[B^9,Y^9]}^2\cr
&+2[B^i,Y^9][Y^i,Y^9]+2[Y^i,B^9][Y^i,Y^9]+{[Y^i,Y^9]}^2)\},\cr
S_{A}=& \int dt Tr \{-\em6 {(\p A)}^2-\em6
{[A,B^9]}^2-{[A,B^i]}^2\cr &+4i\p B^i[A,Y^i]+4i\em6 \p B^9[A,Y^9]-4Q
i\em6 B^9[A,Y^9]\cr & +2i\p Y^i [A,Y^i]-2[A,B^i][A,Y^i]-{[A,Y^i]}^2
\cr &+2i\em6 \p Y^9 [A,Y^9]-2\em6[A,B^9][A,Y^9]-\em6
{[A,Y^9]}^2\}.}}

Since we are considering two D0-branes, the Yang-Mills fields are
just $2\times 2$ matrix. We will choose the background to be
diagonal, \eqn\back{B^1={vt\over2}\sigma_{3}, B^2={b\over
2}\sigma_{3}, B^9={c\over 2}\sigma_{3}.} The background for $A$ and
other transverse directions are chosen to be zero. This corresponds
to, in comoving coordinate, two zero-branes moving towards each
other with relative velocity $v$ in the $x^1$ direction, and
transverse separation $b$ in the $x^2$ direction and $c$ in the
$x^9$ direction. Write the matrix in terms of $U(2)$ generators,
\eqn\utwo{\eqalign{ Y^i&=\h ( \yiz {\bf 1}_{2}+ \yia \sigma^a
),\quad Y^9=\h \et( \ynz {\bf 1}_{2}+ \yna \sigma^a),\cr
A&=\h\et(A_{0}{\bf 1}_{2}+A_{a}\sigma^a ),\quad
 \theta=\h ( \theta_{0}{\bf1}_{2}+\theta_{a}\sigma^a ).}}
 where $a=1,2,3$. The 0 components in
this decomposition describe the free motion of the center of mass
and will not be written explicitly in the following. Then up to
quadratic terms,
 the actions for the fluctuations are
\eqn\sy{\eqalign{S_{i}=\h \int dt\{
&\yio(-{\p}^2-b^2\es-v^2t^2\es-c^2)\yio\cr
+&\yit(-{\p}^2-b^2\es-v^2t^2\es-c^2)\yit\cr
+&\yih(-{\p}^2)\yih\},\cr S_{9}=\h \int dt
\{&\yno(-{\p}^2+Q^2-b^2\es-v^2t^2\es-c^2)\yno\cr
+&\ynt(-{\p}^2+Q^2-b^2\es-v^2t^2\es-c^2)\ynt\cr
+&\ynh(-{\p}^2+Q^2)\ynh\},\cr S_{A}=-\h \int dt
\{&\ao(-{\p}^2+Q^2-b^2\es-v^2t^2\es-c^2)\ao\cr
+&\at(-{\p}^2+Q^2-b^2\es-v^2t^2\es-c^2)\at\cr
+&\ah(-{\p}^2+Q^2)\ah\cr +&4v\et(\ao \yot-\at\yoo)-4Q c(\ao \ynt-\at
\yno)\}.}} Define \eqn\dia{\eqalign{Y_2^9&={1\over
\sqrt{2}}(A_1^++A_1^-),\quad A_1={1\over i\sqrt{2}}(A_1^+-A_1^-),\cr
Y_1^9&={1\over \sqrt{2}}(A_2^++A_2^-),\quad A_2={1\over
i\sqrt{2}}(A_2^- -A_2^+).}}

Then the actions for $A$ and $X_9$ become \eqn\apm{\eqalign
{S_{A^+}=\h \int dt \{&\tao(-{\p}^2+Q^2-b^2\es-v^2t^2\es-c^2-i2Q
c)\tao \cr +&\tat(-{\p}^2+Q^2-b^2\es-v^2t^2\es-c^2-i2Q c)\tat \cr
+&\ah(-{\p}^2+Q^2)\ah+i2\sqrt{2}v \et[(\tao-\tyo)
\yot+(\tat-\tyt)\yoo]\}, \cr S_{A^-}=\h \int dt\{&
\tyo(-{\p}^2+Q^2-b^2\es-v^2t^2\es-c^2+i2Q c)\tyo \cr
+&\tyt(-{\p}^2+Q^2-b^2\es-v^2t^2\es-c^2+i2Q c)\tyt \cr
+&\ynh(-{\p}^2+Q^2)\ynh \}.}}

Define new fermionic fields, \eqn\fe{\theta_{+}={1\over
\sqrt{2}}(\theta_{1}+i\theta_{2}), \quad \theta_{-}={1\over
\sqrt{2}}(\theta_{1}-i\theta_{2}).} Then the action is
\eqn\sf{S_{f}=\int dt \theta_-^T(i\p+vt\et\gamma_{1}+b\et \gamma_{2}
+c\gamma_{9}){\theta}_+ +\h{\theta_{3}}^T(i\p){\theta}_{3}.}

 The ghost action is determined
by the infinitesimal gauge transformation of $G$,
\eqn\sg{\eqalign{S_{g}& = \int dt {C_{1}}^*(
-\p^2-b^2\es-v^2t^2\es-c^2)C_{1}+{C_{2}}^*
(-\p^2-b^2\es-v^2t^2\es-c^2)C_2 \cr &+{C_{3}}^*(-\p^2)C_{3}.}}

Before doing any calculation, we can see that the fluctuation action
for $X_3^\mu$ is independent of the separation, and hence has
nothing to do with the interaction of the two zero branes. We will
not consider them in the following.

\newsec{Static case}

First we will analyze the situation when $v=0$. This corresponds to
two zero-branes static in the comoving coordinates. To calculate the
one loop interaction, we need to integrate out the quadratic
fluctuation, which can be written in the form of determinants,
\eqn\deti{\eqalign{&\hbox{det}^{-\h}(-\p^2-b^2\es-c^2),
\quad\quad\quad\quad\quad\quad\quad \hbox{for} \quad {Y_{1,2}}^i,
i=1,\cdots 8,\cr &\hbox{det}^{-\h} (-\p^2+Q^2-b^2\es-c^2-i2Q
c),\quad \hbox{for} \quad \tao,\tat,\cr &\hbox{det}^{-\h}
(-\p^2+Q^2-b^2\es-c^2+i2Q c),\quad \hbox{for} \quad \tyo,\tyt,\cr
&\hbox{det}(-\p^2-b^2\es-c^2),\quad\quad\quad\quad\quad\quad\quad\quad
\hbox{for} \quad C_{1,2}, \cr &\hbox{det}(i\p+b\et
\gamma_{2}+c\gamma_{9}),\quad\quad\quad\quad\quad\quad\quad\quad
\hbox{for} \quad \theta_+ .}} We use Schwinger proper time formalism
to calculate the determinants. For any Hermitian operator $\Delta$,
the determinant is represented by
\eqn\rep{\delta\equiv\ln(\det\Delta)={-\int_{0}}^\infty {ds\over
s}Tr e^{-i\Delta s}.} Thus we need to calculate the heat kernel,
$K(t',t;s)\equiv<t'|e^{-i\Delta s}|t>$. $K(t,t';s)$ satisfies the
differential equation and the boundary condition,
\eqn\ker{\eqalign{i\partial_{s}K(t',t;s)&=\Delta K(t',t;s),\cr
K(t',t;0)&=\delta(t-t').}} For the first determinant in \deti,
$\Delta=-\p^2-b^2\es-c^2$. To solve \ker, we first solve the static
shr\"{o}dinger equation \eqn\shrd{\lambda
y_\lambda(t)=(-\p^2-b^2\es-c^2)y_\lambda(t).} The two linearly
dependent solutions of \shrd\ are Bessel functions
\eqn\slt{J_{\pm\kappa}(x), \quad \hbox{for}\quad \kappa \notin Z,
\quad\hbox{or}\quad J_\kappa (x),\quad
Y_\kappa(x),\quad\hbox{for}\quad \kappa\in Z,} where $x={b\over Q
}e^{Q t} ,\quad -(Q\kappa)^2=\lambda+c^2$. $Y_n(x)$ has singlarity
at $x=0$, and are not in consideration. Since the operator $\Delta$
is hermitian, $\lambda$ is real, and $\kappa$ is either real or pure
imaginary.

Using an integral of Bessel function (eq. 6.574.2 of \gsrz)
\eqn\bssl{\inzer {dx\over x}
J_\nu(x)J_\mu(x)={2\sin\pi({\mu-\nu\over2})\over\pi(\mu+\nu)(\mu-\nu)},}
 an orthonormal basis can be constructed,
 \eqn\fun{\eqalign{&y_\w(t)=\sqrt{Q\w\over2\sinh(\pi\w) }[J_{i\w}(x)+J_{-i\w}(x)],\quad \w > 0,\cr
& f_n(t)\equiv \sqrt{4Q n}J_{2n}(x), \quad n=1,2,\cdots}}

To check the orthogonality, \eqn\oth{\eqalign{
 &\ininf dty_\w(t)y_{\w'}^*(t)={Q \w \over2\sinh(\pi\w)}\inzer
 {dx\over Q x}[J_{i\w+\epsilon}(x)+J_{-i\w+\epsilon}(x)]
 [J_{-i\w'+\epsilon}(x)+J_{i\w'+\epsilon}(x)]\cr
&={\w\over \sinh(\pi\w)}\{{\sinh[{\pi\over 2}(\w+\w')]\over
(\w+\w')} {\epsilon\over \pi[({\w-\w'\over 2})^2+\epsilon^2]}+
{\sinh[{\pi\over 2}(\w-\w')]\over
(\w-\w')}{\epsilon\over\pi[({\w+\w'\over 2})^2+\epsilon^2]}\}\cr
 &=\delta(\w-\w')+\delta(\w+\w')\cr
 &=\delta(\w-\w'),\cr
 &\ininf dtf_n(t)f_m(t)=\delta_{n,m},\cr
 &\ininf dtf_\w(t)f_n( t)\propto\sin(n\pi)=0.
  }}We have deformed $\pm i\w$ by a small real
part, $\pm i\w \rightarrow \pm i\w+\epsilon$ and used the identity
$\delta(z)=\lim_{\epsilon\rightarrow 0}{\epsilon \over
\pi(z^2+\epsilon^2)}$. In the fourth line $\w>0$ is taken into
account.

To check the completeness, we will need to prove that all
$J_\kappa(x), \Re\kappa>0, \kappa\neq 2n, n\in Z_+ $ can be expanded
in the basis. Define
\eqn\fouri{\eqalign{&\tilde{J}_\kappa(\w)=\inzer dt y^*_\w
(t)J_\kappa(x),\quad \tilde{J}_\kappa^n=\inzer dt
f_n(t)J_\kappa(x),\cr &\tilde{\tilde{J}}_\kappa(x)=\inzer d\w
\tilde{J}_\kappa(\w)y_\w(t)+\sum^\infty_{n=1}\tilde{J}_\kappa^n
f_n(t) .}} Using \bssl, one finds
that\eqn\frc{\tilde{J}_\kappa(\w)=\sqrt{\w\over2Q\sinh(\pi\w)}{4
\cosh({\pi\w\over2})\sin({\pi\kappa\over2})\over\pi(\kappa^2+\w^2)},\quad\tilde{J}_\kappa^n=4\sqrt{n\over
Q}{(-1)^n\sin(\pi{\kappa\over  2})\over\pi(\kappa^2-4n^2)}.} Hence,
\eqn\cmplt{\inzer d\w \tilde{J}_\kappa(\w)y_\w(t) =\ininf
d\w{\sin({\pi\kappa\over 2})\w
J_{i\w+\epsilon}(x)\over\pi\sinh({\pi\w\over 2})( \kappa^2+\w^2)}.}
The large order behavior of the Bessel function is
\eqn\larg{J_\mu(x)\sim e^{\mu+\mu\ln{x\over 2}-(\mu+\h)\ln
\mu}.}Then the integral \cmplt\ can be evaluated by closing the
contour in the lower half plane. Simple poles are at $\w=-2ni,
-i\kappa$. \eqn\invf{\eqalign{\inzer d\w
\tilde{J}_\kappa(\w)y_\w(t)&=J_\kappa(x)-{8\over\pi}\sin({\pi\kappa\over
2})\sum^\infty_{n=1}(-1)^n{n\over \kappa^2-4n^2}J_{2n}(x)\cr
&=J_\kappa(x)-\sum^\infty_{n=1}\tilde{J}_\kappa^n f_n(t) ,
}}Therefore, \eqn\cprf{\tilde{\tilde{J}}_\kappa(x)=J_\kappa(x).}
When $\kappa=\pm i\w+\epsilon $, the above equations still hold.
Then the other linear combination of $J_{\pm\w}(x),\quad
J_{i\w}(x)-J_{-i\w}(x)$ can be also expanded in terms of the basis
\fun, and so are not included in the basis. In fact, we have shown
that any normalizable eigenfunction can be expanded in terms of this
basis, which is enough to guarantee that the basis is complete. (The
completeness of this set of the eigenfunctions was discussed
previously in \stefan.)

 Then the heat kernel can be expanded in
terms the orthonormal basis,
\eqn\kernel{K(t',t;s)={\int_{0}}^{\infty}d\w y_\w(t')^*
y_\w(t)e^{-i[{(Q\w )}^2-c^2]
s}+\sum^\infty_{n=1}f_n(t')f_n(t)e^{i[(2Q n)^2+c^2]s}.}

\subsec{The bosonic effective potential}

Having found out the heat kernel, it is straight forward to write
down the determinant explicitly, by $\delta_i=-\inzer {ds\over
s}\ininf dt K(t,t;s)$. The trace in \rep\ is now an integral over
$t$. In order to compare with the result obtained on the
supergravity side \mls, we need to compactify the 9-direction and
smear the result over the circle. This is equivalent to sum the
images in the covering space and then average over the compactified
circle. On the matrix theory side, we need to calculate the one loop
effective potential of two D0-branes separated also by $c$ in the
$x^9$ direction, integrate over $c$ and then divide by $2\pi R'$.
This procedure is expected to give us the result that is to be
compared with our earlier result in \mls. Here $R'$ is the radius of
$X^9$. Now there are altogether four integrals in our calculation of
determinant, the integral over $t, s, \w,$ and $c$. We can first do
the the integral over $c$. Then the smeared determinant becomes
\eqn\avrg{\eqalign{\delta_{i}={-1\over 2 \sqrt{-\pi i}R'} \inzer
{ds\over s^{3\over2}} \ininf dt\{&\ininf d\w {Q\w\over2\sinh (\pi
\w)}J_{i\w}(x)[J_{i\w}(x)+J_{-i\w}(x)] e^{-i{(Q\w)}^2s}\cr
&+\sum^\infty_{n=1}4Q n J^2_{2n}(x)e^{i(2Q n)^2s}\} \cr={-1\over 2
\sqrt{-\pi i}R'} \inzer {ds\over s^{3\over2}} \ininf dt\{&\ininf d\w
{Q\w\over2\sinh (\pi
\w)}J_{i\w}(x)[J_{i\w}(x)+J_{-i\w}(x)]\cr\times[1-i(Q\w)^2s+\cdots]
+& \sum^\infty_{n=1}4Q n J^2_{2n}(x)[1+i(2Qn)^2s+\cdots]\}.}} Here
we have extended the integral range of $\w$ from $(0, \infty)$ to
$(-\infty, \infty)$. We use the notation $\pm i=e^{\pm\pi i\over2}$,
$\sqrt{\pm i}=e^{\pm\pf i}$, and $\ln(i)={\pi i\over 2} $. We have
rewritten the exponential in the form of power series. Using the
large order behavior of Bessel function, we have
\eqn\lbsl{\eqalign{{\w^{2n+1}\over\sinh(\pi\w)}J_{i\w}(x)J_{-i\w}(x)&\sim
{\w^{2n+1}\over\sinh(\pi\w)}\exp[-2i\w\ln i-\ln \w]=
{\w^{2n}e^{\pi\w}\over \sinh(\pi\w)},\cr
 {\w^{2n+1}\over\sinh(\pi\w)}J^2_{i\w}(x)&\sim
{\w^{2n+1}\over\sinh(\pi\w)}\exp[2i\w+2i\w\ln
{x\over2}-2i\w\ln\w-2i\w\ln i-\ln \w]\cr &={\w^{2n}\exp[2i\w+2i\w\ln
{x\over2}-2i\w\ln\w+\pi\w]\over\sinh(\pi\w)}.}} Close the contour in
the lower half plane, we can see that the integral of each term
proportional to $J^2_{i\w}(x)$ at the infinity is zero. Because of
the third line of \lbsl, we will meet a divergence at infinity in
each term proportional to $J_{i\w}(x)J_{-i\w}(x)$. Note that this
divergence is independent of the $x$, and therefore can be
subtracted. Then \eqn\omg{\eqalign{&\ininf d\w
{Q\w\over2\sinh(\pi\w)}J_{i\w}(x)[J_{i\w}(x)+J_{-i\w}(x)][1-i(Q
\w)^2s+\cdots],\cr &=-4Q n J^2_{2n}(x)[1+i(2Q n)^2s+\cdots].}}  The
above just cancels with the summation in \avrg\ term by term.
Although we are not sure about the convergence of the expansion, the
exact cancelation of each term between the integral \omg\ and the
summation in \avrg\ has show that $\delta_i=0$ up to a physical
irrelevant constant. So the bosons coming from the $i$ directions
give no contribution to the effective potential.


For the second and the third determinants in \deti, The heat kernel
becomes \eqn\ka{\eqalign{K_+(t',t;s)&=\int_0^\infty d\w
y^*_\w(x')y_\w(x)e^{-i[{(Q\w )}^2-(c+iQ)^2] s}
+\sum^\infty_{n=1}f_n(t')f_n(t)e^{i[(2Q n)^2+(c+iQ)^2]s},\cr
K_-(t',t;s)&={\int_{0}}^{\infty}d\w y^*_\w(x') y_\w(x) e^{-i[(Q\w
)^2-(c-iQ)^2] s}+\sum^\infty_{n=1}f_n(t')f_n(t)e^{i[(2Q
n)^2+(c-iQ)^2]s}.}} Then take the same procedure as in eqs. \avrg,
\lbsl, and \omg, we will find that the bosons coming from the gauge
field and $X^9$ give no contribution to the effective potential,
either.


The ghost determinant is the same with that of $X^i$, and hence give
the same result except for a minus sign.

In a word, we find that there is no static potential coming from the
bosons.

\subsec{The fermionic effective potential}

In the fermionic sector, there are 16 degrees of freedom for each
$SU(2)$ index. Since there are only three gamma matrix relevant
here, we can choose a basis to make the gamma matrix and the field
block diagonal, \eqn\gama{\gamma_1=\sigma_2\otimes {\bf 1}_8,\quad
\gamma_2=\sigma_3\otimes {\bf 1}_8, \quad \gamma_9=\sigma_1\otimes
{\bf 1}_8.} Define $$
K_{\alpha\beta}(t',t;s)=<t'|\exp(-i\Delta_f(\hat{t})
s)|t>_{\alpha\beta}, $$
 where $\Delta_f=i\p+b\et \gamma_{2}+c\gamma_{9},\quad \alpha ,\beta=1,2$
 label the two $8\times 8$ block matrix.
 Then $\kab$ satisfies the following differential equation and initial condition
\eqn\keqn{i\partial_s\kab=(\Delta_f(\hat{t}))_{\alpha\rho}K_{\rho\beta}(t',t;s)
,\quad K_{\alpha\beta}(t',t;0)=\delta_{\alpha\beta}\delta(t'-t).} To
find the solution, we write
$$\left(\matrix{K_{1\beta}(t',t;s)\cr K_{2\beta}(t',t;s)}\right)
=\ininf d\lambda K_\beta(t',t;\lambda)e^{-i\lambda s}.$$ In the
following, we just write $K_\beta$ for short. Then from \keqn,
$K_\beta$ satisfy the following differential equations
\eqn\eign{\eqalign{ &(i\p+b\et-\lambda)K_{1\beta}+cK_{2\beta}=0,\cr
&(i\p-b\et-\lambda)K_{2\beta}+cK_{1\beta}=0. }} These equations are
equivalent to \eqn\egn{\eqalign{ &cK_{2\beta} =-( i
\p+b\et-\lambda)K_{1\beta},\cr &(\p^2+b^2 \es+c^2-\lambda^2+2
i\lambda\p
 -i Q b \et) K_{1\beta}=0.}} Denote $\lambda/(Q)$ by $\w$,
and $c/Q$ by $ c'$. Take the ansatz
$K_{1\beta}=f(w,c',t')\psi(\w,c',t)$. Then $f(\w,c',t')$ factorize
and the equation for $\psi(\w,c',t)$ has two linearly independent
solutions, $x^{(-1/2-i\w)}M_{1/2,ic'}(-2ix)$ and
$x^{(-1/2-i\w)}M_{1/2,-ic'}(-2ix)$. Where $M_{\lambda,\mu}(x)$ is a
Whittaker function. A general solution for \egn\ is
\eqn\sol{\eqalign{K_\beta=&
f(\w,c',t')x^{(-1/2-i\w)}\left(\matrix{M_{1/2,ic'}(-2ix)\cr
M_{-1/2,ic'}(-2ix)}\right)\cr &+g(\w,c',t')
x^{(-1/2-i\w)}\left(\matrix{M_{1/2,-ic'}(-2ix)\cr -
M_{-1/2,-ic'}(-2ix)}\right),}} where $f(\w,c',t')$ and $g(\w,c',t')$
are chosen to satisfy the initial condition. Then
\eqn\initial{\eqalign{K_{\alpha 1}(t',t;s)= &\ininf d\w Q ({x\over
x'})^{-i\w} {\exp (-iQ \w s)\over-4i\sqrt{xx'}}[
M_{-1/2,-ic'}(-2ix') \left(\matrix{M_{1/2,ic'}(-2ix)\cr
M_{-1/2,ic'}(-2ix)}\right) \cr  &+
M_{-1/2,ic'}(-2ix')\left(\matrix{M_{1/2,-ic'}(-2ix)\cr -
M_{-1/2,-ic'}(-2ix)}\right)] ,\cr
 K_{\alpha
2}(t',t;s) = &\ininf d\w Q ({x\over x'})^{-i\w} {\exp (-iQ \w
s)\over-4i\sqrt{xx'}}[M_{1/2,-ic'}(-2ix')
\left(\matrix{M_{1/2,ic'}(-2ix)\cr  M_{-1/2,ic'}(-2ix)}\right) \cr &
-M_{1/2,ic'}(-2ix')\left(\matrix{M_{1/2,-ic'}(-2ix)\cr-
M_{-1/2,-ic'}(-2ix)}\right)] .}} When $s=0$, $\w$ can be integrated
out and gives $\delta(t-t')$. Then
$$K_{\alpha\beta}(t',t;0)=\delta_{\alpha\beta}\delta(t-t'){D(x,x',c')\over
-4i\sqrt{xx'}},$$ where \eqn\indt{D(x,x',c')\equiv
M_{1/2,ic'}(-2ix)M_{-1/2,-ic'}(-2ix')+M_{1/2,-ic'}(-2ix)M_{-1/2,ic'}(-2ix').
} In appendix B, we will prove that $D(x,x,c')=-4ix$. So \initial\
satisfies the initial condition in \keqn.

Finally, \eqn\phsaf{\eqalign{\delta_f&=-\ininf {dc\over 2\pi R'}
\int_0^\infty {ds\over s} tr K_{\alpha,\beta}(t,t';s)\cr &=-\ininf
dt\ininf{dc\over 2\pi R'} \int_0^\infty {ds\over s} \ininf d\w Q
 \exp[-iQ\w s]\cr & \times D^{-1}
[M_{1/2,ic'}(-2ix)M_{-1/2,-ic'}(-2ix)+M_{1/2,-ic'}(-2ix)M_{-1/2,ic'}(-2ix)]\cr
&=-\ininf dt\ininf{dc\over 2\pi R'} \int_0^\infty {ds\over s} \ininf
d\w Q \exp[-iQ\w s]}} The $tr$ in the first line means a trace of
the $2\times2$ matrix and an integral over $t$. The integral is
divergent but is independent of $b$ and $\alpha$. In fact, this is
just the phase shift generated by a free operator $i{d\over dt}$.
Regularize the phase shift by subtracting $-\inzer{ds\over s}tr
e^{-iH_0s}$, $H_0$ is the Lagrange for free fermions. We can see
that the fermions give no contribution to the effective potential.

Then we can draw the conclusion that there is no static effective
potential.

\newsec{Effective interaction at the order $v^2$}

Now we are going to investigate the case when there is a small
relative velocity between the zero branes. Since it is difficult to
compute the determinants directly, we will perturbatively expand
around $v=0$.

From \sy, \apm, \sf and \sg, we can see that the only possible terms
term odd in $v$ in the perturbation series come from the the
fermionic action \sf. These terms vanish because the trace of odd
number of Gamma matrix is zero.We shall in this section calculate
various $v^2$ terms.

Denote the $v^2$ terms coming from the first order bosonic
contribution by $b_{1}$, the second order bosonic contribution by
$b_{2}$,
 the first order ghost contribution by $g_{1}$, and the
second order fermionic contribution by $f_{2}$. Then,
\eqn\vt{\eqalign{b_1=&-i \ininf {dc\over 2\pi R'} \ininf dt \h
v^2t^2 \es [\langle Y_1^i(t)Y_1^i(t)\rangle+\langle
Y_2^i(t)Y_2^i(t)\rangle,\cr &+\langle
A_1^+(t)A_1^+(t)\rangle+\langle A_2^+(t)A_2^+(t)\rangle +\langle
A_1^-(t)A_1^-(t)\rangle+\langle A_2^-(t)A_2^-(t)\rangle ],\cr g_1=&i
\ininf {dc\over 2\pi R'} \ininf dt v^2t^2 \es [\langle
C_1(t)C_1^*(t)\rangle+\langle C_2(t)C_2^*(t)\rangle],\cr
b_{2}=&\h\ininf {dc\over 2\pi R'} \ininf dt_{1}\ininf dt_{2}
(v\sqrt{2})^2 e^{Q t_1}e^{Q t_2}\cr &\times\{ [<
\tao(t_1)\tao(t_2)>+< \tyt(t_1)\tyt(t_2)>]< \yot(t_1)\yot(t_2)>\cr
&+[< \tyo(t_1)\tyo(t_2)>+< \tat(t_1)\tat(t_2)>]<
\yoo(t_1)\yoo(t_2)>\},\cr f_{2}=&{1\over 2} \ininf {dc\over 2\pi
R'}\ininf dt_{1}\ininf dt_{2} v^2 t_{1}t_{2} e^{Q t_{1}}e^{Q
t_{2}}\cr &\times \Tr[\gamma_1 G_f(t_1,t_2)\gamma_1 G_f(t_2,t_1)].}}

Since we are only interested in the effective potential, we do not
have to do all the integrals in the second order contributions.
Define $t_1=t+\h\tau,\quad t_2=t-\h\tau$, integrate out $\tau$ and
$c$, we are left with an integral of $t$, which combined with the
first order perturbation, will give the effective potential to
$v^2$ order. In Appendix A, we will show how this procedure is
carried out when the background is flat. We hope that this
procedure also goes through here, as we shall see, there is a
problem arising at this order.

The propagator $<Y_a^i(t_2)Y_a^i(t_1)>\equiv G_{i}(t_2,t_1)$,
satisfies the differential equation
\eqn\bprp{(-\partial_{t_1}^2-b^2e^{2Qt_1}-c^2)G_{i}(t_2,t_1)=i\delta(t_1-t_2),}and
is related to the heat kernel by $G_{i}(t_2,t_1)=-\inzer ds
K(t_2,t_1;s).$ \eqn\prpgt{\eqalign{G_{i}(t_2,t_1)=&-\inzer ds
K(t_2,t_1;s)\cr =&i\ininf d\w {Q \w \over
2\sinh(\pi\w)}J_{i\w}(x_1)[J_{-iw}(x_2)+J_{i\w}(x_2)] / [(Q
\w)^2-c^2-i\epsilon ]\cr &+i\sum^\infty_{n=1}4Q n
J_{2n}(x_1)J_{2n}(x_2)/[-(2Qn)^2-c^2-i\epsilon]\cr =&
-\theta(t_1-t_2){\pi\over2Q \sinh(\pi c')}J_{- ic'}(x_2)[J_{
ic'}(x_1)+J_{-ic'}(x_1)]\cr &-\theta(t_2-t_1){\pi\over2Q \sinh(\pi
c')}J_{-ic'}(x_1)[J_{ ic'}(x_2)+J_{-ic'}(x_2)].}} From the second
line of \prpgt\ to the forth line, we have integrated $\w$ by
contour integral, and assumed $c'>0$. When $c'<0$, just replace $c'$
with $-c'$. Using the asymptotical behavior of Bessel function at
large order \larg, we see that $J_{i\w}(x_1)J_{-i\w}(x_2)\sim
e^{iQ\w(t_1-t_2)+\pi\w-\ln\w}$. When $t_1-t_2>0$, we should close
the contour in the upper half plane, and otherwise the lower half
plane. The poles at $\pm 2ni$ cancels with the sum in the third line
of \prpgt. So only the poles at $\pm(c+i\epsilon) $ contribute to
the propagator. It is difficult to obtain a compact result. We will
take a $b\rightarrow \infty$ limit to obtain the asymptotic
behavior. Or equivalently, we let $b/(Q)$ to be of order 1, and let
$t\rightarrow \infty $ .
\eqn\gtp{G_i(t_2,t_1)\sim-\theta(t_1-t_2){\cos(x_2+ \h i\pi
c'-{\pi\over4})\cos(x_1-\pf)\over
Q\sqrt{x_1x_2}\sinh\pic}+(t_1\leftrightarrow t_2).}
 We use $G_i(t,t)$ to calculate $b_1$, which
 becomes \eqn\apro{G_i(t,t)\sim-{[\sin(2x)+1]
 \coth\pic\over 2Q x}-{i\cos(2x)\over 2Qx}.}

 Define \eqn\apm{\eqalign{G_+(1,2)&=<
\tao(t_1)\tao(t_2)>=< \tat(t_1)\tat(t_2)>,\cr G_-(1,2)&=<
\tyo(t_1)\tyo(t_2)>=< \tyt(t_1)\tyt(t_2)>.}} They satisfy the
following differential equations
 \eqn\gadiff{\eqalign{(-\partial^2_{t_1}-b^2e^{2Qt_1}-(c+iQ)^2
 )G_+(t_2,t_1)&=i\delta(t_1-t_2),\cr
 (-\partial^2_{t_1}-b^2e^{2Qt_1}-(c-iQ)^2 )G_-(t_2,t_1)&=i\delta(t_1-t_2)
 .}}
The solution $G_+$ can also be obtained from the heat kernel.
\eqn\gasol{\eqalign{G_+(t_2,t_1)=&-\inzer ds K_+(t_2,t_1;s)\cr
=&\int_0^\infty d\w {iQ\w[J_{i\w}(x_1)+J_{-i\w}(x_1)]
[J_{i\w}(x_2)+J_{-i\w}(x_2)]\over2\sinh(\pi\w)[(Q\w
)^2-(c+iQ)^2]}\cr &+i\sum^\infty_{n=1}{4Q n
J_{2n}(x_1)J_{2n}(x_2)\over -(2Q  n)^2-(c+iQ)^2}\cr
=&-{\theta(t_1-t_2)\pi\over2Q \sinh[\pi
(c'+i)]}J_{-i(c'+i)}(x_2)[J_{i(c'+i)}(x_1)+J_{-i(c'+i)}(x_1)]\cr
&-{\theta(t_2-t_1)\pi\over2Q \sinh[\pi
(c'+i)]}J_{-i(c'+i)}(x_1)[J_{i(c'+i)}(x_2)+J_{-i(c'+i)}(x_2)]\cr
\sim& i\theta(t_1-t_2){\cos[x_2+\h
i\pi(c'+i)-{\pi\over4}]\cos(x_1-\pf)\over
Q\sqrt{x_1x_2}\cosh\pic}+(t_1\leftrightarrow t_2)
 ,}}and \eqn\gatt{G_+(t,t)\sim-{[\sin(2x)+1]
 \tanh\pic\over 2Q x}-{i\cos(2x)\over 2Qx}.}

 To
get $G_- $, just replace $c'$ by $-c'$  in \gasol.

Then the first order contribution adds up to
\eqn\vone{\eqalign{-iV_1&=-i\ininf {dc\over 2\pi R'}
v^2t^2\es[6G_i(t,t)+G_+(t,t)+G_-(t,t)]\cr
&\sim{i3v^2t^2\et[1+\sin(2x)]\over b}\inzer {Qdc'\over\pi R'}
\coth\pic-{4v^2t^2\et\cos(2x)\over b}\inzer {Qdc'\over \pi R'}\cr
&\sim {i3v^2t^2\et\over b}\inzer {Qdc'\over\pi R'}\coth\pic.}} In
the last step, we have omitted the trigonometric functions because
they are periodic functions and fluctuate violently at large
argument. The integral of $c'$ seems to give a divergence, but this
is just caused by our using the asymptotic expansion of the bessel
function. That step hides the depression of the large $c'$. In fact,
from the large order behavior \lbsl, we can see that there is indeed
no divergence in the $c'$. Hereafter, we can just put the this
divergence aside.

To compute $b_2$, we again need to take the limit $|x_1|\gg 1$ and
$|x_2|\gg 1$. This is equivalent to $t_1\gg 1$ and $t_2\gg 1$. In
this case, we use the asymptotic expansion of Bessel function when
$x_1$ and $x_2$ are large in \prpgt\ and \gasol.  Multiply \prpgt\
and \gasol, and integrate out $\tau\equiv t_1-t_2$, we will get the
dependence on $x={be^{Q t}\over Q}, \quad t=\h(t_1+t_2).$  The
relevant integral is \eqn\intau  {\eqalign{-i V_{b2}= &\inzer
{dc\over\pi R'}2 v^2e^{2Q t}\ininf d\tau
G_i(t_1,t_2)[G_+(t_2,t_1)+G_-(t_2,t_1)]\cr \sim&\inzer {dc\over\pi
R'} \inzer d\tau{\theta(t_1-t_2)v^2\over b^2}\{i\coth\pic[\cos
(2x_2)+\h\sin( 2x_1+2x_2)\cr
&+\h\sin(2x_1-2x_2)]-[1-\sin(2x_2)+\sin(2x_1)\cr
&-\h\cos(2x_1-2x_2)+\h\cos(2x_1+2x_2)]\} +(t_1\leftrightarrow
t_2)\cr  \sim &\inzer {dc'\over\pi R'}{v^2\over
b^2}\{i\pi\coth\pic[I_0(4x)-{\bf L}_0(4x)]-\ininf
Qd\tau+2K_0(4x)\}.}} Considering $x_1\gg 1,$ and $x_2\gg 1$,  The
trigonometric functions with argument $x_1$, $x_2$, and $2x_1+2x_2$
fluctuate quickly in the $x\gg 1$ limit, they average to zero and
hence can be omitted. The term $\ininf Qd\tau$ seems to be divergent.
However, remember the limit we are taking here, $t_1, t_2\gg1$,
and $t$ fixed, so the range of both $t_1$ and $t_2$ is proportional to $t$.
Then the range of $\tau= t_1-t_2 $ is also proportional to $t$, and the term
proportional to $\ininf Qd\tau$ is finite
and increases with $t$.

To calculate $f_2$, we will need the fermionic propagator, defined
by $G_{\alpha,\beta}(t_1,t_2)\equiv \langle
T\theta_+(t_1)\theta_-^T(t_2) \rangle_{\alpha\beta}.$ It satisfies
the following differential equation,
\eqn\diff{[i\partial_{t_1}+be^{Qt_1}
\gamma_{2}+c\gamma_{9}]G_{\alpha,\beta}(t_1,t_2)=-i\delta_{\alpha,\beta}\delta(t_1-t_2).}
The propagator is related to the heat kernel
$K_{\alpha\beta}(t_2,t_1;s)$ roughly by
$G_{\alpha,\beta}(t_1,t_2)=\inzer ds K_{\alpha\beta}(t_2,t_1;s).$
But there is some subtly in determining the time ordering in each
term. This is related to the boundary conditions. We are not going
to solve the problem in this way. Instead, we take the $b\rightarrow
0$ limit. The limiting case will be the propagator for massive
fermions, which will be analyzed in Appendix A. When $b\rightarrow
0$, the argument of the Whittaker function is small. We will have
\eqn\aswht{M_{\lambda,\mu}(z)\sim z^{\mu+\h}e^{-{z\over 2}} .} We
determine the time ordered propagator by comparing its small $b$
limit with the propagator of a massive fermion in Appendix A ,
consequently
 \eqn\gr{\eqalign{G_{11}(t_1,t_2) =(4i\sqrt{x_1x_2})^{-1}
[&\theta(t_1-t_2)M_{-1/2,-ic'}(-2ix_2)M_{1/2,ic'}(-2ix_1)\cr
-&\theta(t_2-t_1)M_{-1/2,ic'}(-2ix_2)M_{1/2,-ic'}(-2ix_1)]\cr
G_{22}(t_1,t_2)=(4i\sqrt{x_1x_2})^{-1}
[&\theta(t_1-t_2)M_{1/2,-ic'}(-2ix_2)M_{-1/2,ic'}(-2ix_1)\cr
 -&\theta(t_2-t_1)M_{1/2,ic'}(-2ix_2)M_{-1/2,-ic'}(-2ix_1)]\cr
G_{12}(t_1,t_2)=(4i\sqrt{x_1x_2})^{-1}
[&\theta(t_1-t_2)M_{1/2,-ic'}(-2ix_2)M_{1/2,ic'}(-2ix_1)\cr
+&\theta(t_2-t_1)M_{1/2,ic'}(-2ix_2)M_{1/2,-ic'}(-2ix_1)]\cr
G_{21}(t_1,t_2)=(4i\sqrt{x_1x_2})^{-1}
[&\theta(t_1-t_2)M_{-1/2,-ic'}(-2ix_2)M_{-1/2,ic'}(-2ix_1)\cr
+&\theta(t_2-t_1)M_{-1/2,ic'}(-2ix_2)M_{-1/2,-ic'}(-2ix_1)]. }}
Again let $b/(Q)$ to be of order 1, and take the $t_{1,2}\gg 1$
limit also, we will get a finite integral with respect to both
$\tau$. In the following, we will need to use the asymptotical
expansion of Whittaker function at large argument.
\eqn\whias{\eqalign {& M_{\h,\pm ic'}(-2ix)\sim {\Gamma(\pm
2ic'+1)\over\Gamma(\pm ic'+1)}\exp(\mp\pi c'+ix)\sqrt{-2ix}\cr
&M_{-\h,\pm ic'}(-2ix)\sim{\Gamma(\pm 2ic'+1)\over\Gamma(\pm
ic'+1)}\exp(-ix)\sqrt{-2ix}.}} Then

 \eqn\intgr{\eqalign{\Tr[\gamma_1 G_f(t_1,t_2)\gamma_1
G_f(t_2,t_1)]=
[&G_{11}(t_2,t_1)G_{22}(t_1,t_2)+G_{22}(t_2,t_1)G_{11}(t_1,t_2)\cr
-&G_{12}(t_1,t_2)G_{12}(t_2,t_1)-G_{21}(t_1,t_2)G_{21}(t_2,t_1)]\cr
=&(8x_1x_2)^{-1}\theta(t_1-t_2)[M_{-1/2,-ic'}^2(-2ix_2)M_{1/2,ic'}^2(-2ix_1)\cr
&+M_{1/2,-ic'}^2(-2ix_2)M_{-1/2,ic'}^2(-2ix_1)]+(t_1\leftrightarrow
t_2)\cr \sim &-{\cosh(2ix_2-2ix_1+2\pi c')\over \cosh^2(\pi c')}
 }}
The fermionic contribution to the effective potential is thus
\eqn\feras{\eqalign{-iV_f\equiv & \inzer {dc\over \pi R'} \ininf
d\tau 8v^2e^{2Q t}(t^2-{\tau^2\over 4})\Tr[\gamma_1
G_f(t_1,t_2)\gamma_1 G_f(t_2,t_1)]\cr \sim &\inzer {8dc'\over \pi
R'}\{ -{v^2\es\cosh(2\pi c')\over \cosh^2(\pi
c')}K_0(x)[4t^2-{\pi^2\over Q^2}]\cr &-{i2\pi v^2t^2\es\sinh(2\pi
c')\over \cosh^2(\pi c')}[I_0(4x)-{\bf L}_0(4x)]\}}}

From \vone, \intau, and \feras\ we see that the effective potential
proportional to $v^2$ does not vanish. The late time potential
contains both a real part and an imaginary part. In the following,
when we use ``proportional to", we mean that we ignore some
numerical coefficient, including the the integral of $c'$.  The
leading real part comes from the bosons, \vone, proportional to
$-Qv^2t^2e^{Qt}\over b$. It increases with $t$. The leading imaginary
part also comes from the bosonic part, \intau, proportional to
$-i{v^2\over b^2}\ininf Qd\tau$. This term is finite and increases as $t$ as we have explained
 following eq.\intau.   We may also pay
attention to the subleading terms, which are finite, and may have
some physical significance. The subleading imaginary part comes from
the fermionic contribution, which is proportional to $-iv^2t^2 e^{2Q
t}K_0(4x)\sim {-iv^2t^2 e^{2Q t}\over \sqrt{b e^{Q t}\over
Q}}e^{-4e^{Q t}\over Q}$.
 The subleading real contribution comes
also from the fermionic contribution, which is proportional to
$v^2t^2 e^{2Q t}[I_0(4x)-{\bf L}_0(4x)]\sim {v^2t^2 e^{2Q t}\over
\sqrt{b e^{Q t}\over Q}}e^{-4e^{Q t}\over Q}$. Both the  real part
and the imaginary part of the effective potential are proportional
to positive power of $Q$, so when $Q \rightarrow 0,$ both vanish.
The subleading effective potential also vanish as $t\rightarrow
\infty$.

\newsec{Conclusion and Discussions}
 We study the effective potential between two D0-branes in a time-dependent
 matrix theory at the one loop level. When the two D0-branes have no relative
 motion in the comoving coordinates, we find that there is no
 effective potential. This result is expected if there is
 supersymmetry, thanks to the cancelation between bosons and fermions.
 What is surprising is that there is no supersymmetry in our case.
 The bosonic and fermionic phase shifts are both divergent but do not dependent on the
 physical parameter, the seperation $b$. So upon suitable
 regularization, they are both zero.

 When we consider the case when $v\neq 0$, the exact form of the
 effective potential is not calculated because of the integrand is
 too complicated. Too see that the potential is non-trivial, we
 examined the behavior of the potential in later times. The
 $v^2$ corrections do not cancel in one loop calculation. Moreover, there exists an
 imaginary part in addition to a real part. This
result seems to contradict with our supergravity calculation. When
we compactify the $X^9$ direction, we get a type IIA string theory
with string coupling constant $g_s=e^{-Q t}$, and the effective 10
dimensional gravitational constant is $\kappa^2\propto g_s^2=e^{-2Q
t}$. Supergravity loop expansion is in terms of gravitational
constant. But we see no sign of this expansion in matrix
calculation. Furthermore, the imaginary part of the effective
potential may imply an instability of the 2 D0-brane system. As the
two D0-branes move apart in the comoving coordinates, certain modes
in the two D0-brane system become tachyonic, and the imaginary part
just signals creation of these modes.

\noindent {\bf Acknowledgments}

We are grateful to Cheng Guo, Qing Guo Huang, Tao Wang  and Peng
Zhang for useful discussions. This work was supported by a grant
from CNSF. We are also grateful to the referee for pointing out the
problem of completeness in our previous version.
\appendix{A}{Perturbation in flat background }

When the background is flat, BFSS matrix model has been tested to
two loops. Here we will use our perturbation method to repeat the
result to one loop order. Set $b=0$ and $Q=0$, we just return to the
situation investigated by \becker. The $v=0$ case is similar. The
determinants we are going to compute becomes
\eqn\detbnot{\eqalign{&\hbox{det}^{10}(-\p^2-c^2) \quad \hbox{for}
\quad Y_{1,2}^\mu, \mu=1,\cdots 9\quad \hbox{and} \quad A_{1,2},\cr
&\hbox{det}^{-2}(-\p^2-c^2)\quad \hbox{for} \quad C_{1,2} ,\cr
&\hbox{det}^{-8}(i\p+c\gamma_{9})\quad \hbox{for} \quad \theta_+.}}
The propagators are $G_b(t,t')=-{1\over 2c}e^{ic|t-t'|}$ for all the
bosons and the ghosts. For the fermions,
 \eqn\gbbot{\eqalign{
G_{11}(t,t')&=G_{22}(t,t')=-\epsilon(t-t')e^{ic|t-t'|},\cr
G_{12}(t,t')&=G_{21}(t,t')=-\h e^{ic|t-t'|}, }} where
$\epsilon(t-t')=\h[\theta(t-t')-\theta(t'-t)].$ For the $v\neq 0$
case, we need \eqn\vtbnot{\eqalign{b_1&=-i \ininf dt \h v^2t^2
[\langle Y_1^\mu(t)Y_1^\mu(t)\rangle +\langle
Y_2^\mu(t)Y_2^\mu(t)\rangle\cr &+\langle A_1(t)A_1(t)\rangle
+\langle A_2(t)A_2(t)\rangle ],\cr g_1&=i   \ininf dt  v^2t^2
[\langle C_1(t)C_1^*(t)\rangle +\langle C_2(t)C_2^*(t)\rangle ],\cr
b_{2}&=- \ininf \ininf dt_{1}dt_{2} v^2 \{ \langle \ao(t_1)\ao(t_2)
\rangle \langle\yot(t_1)\yot(t_2)\rangle\cr &+\langle
\at(t_1)\at(t_2)\rangle\langle \yoo(t_1)\yoo(t_2)\rangle \},\cr
f_{2}&={1\over 2} \ininf \ininf dt_{1}dt_{2} v^2 t_1t_2 \times
8Tr[\gamma_1 G_f(t_1,t_2)\gamma_1 G_f(t_2,t_1)].}} In order to get
the effective action, we do not need to perform all the integrals.
Define $t={t_1+t_2\over 2 }\quad \tau=t_1-t_2$, integrate out
$\tau$, and sum over all terms above, we will get the effective
potential before the smearing: \eqn\vtwo{\eqalign{b_1&=i \ininf dt
v^2t^2 {5\over c},\cr g_1&=-i   \ininf dt v^2t^2 {1\over  c},\cr
b_2&=-i\ininf dt v^2 {1\over 2c^3},\cr f_2&=-i\ininf dt
v^2({4t^2\over c}-{1\over 2c^3 }).}} The various factors comes from
the counting of degree of freedom. They sum up to zero. So there is
no $v^2$ term in the effective action.

\appendix{B}{The proof of an identity}

Here we will give the proof of the following identity
\eqn\absid{D(c,z)\equiv
M_{1/2,ic}(z)M_{-1/2,-ic}(z)+M_{1/2,-ic}(z)M_{-1/2,ic}(z) =2z,}
where $z$ is pure imaginary. In the following, we will treat
$D(c,z)$ as a function of $c$, and view $z$ as a parameter. Using
the steepest descendent method, we can get the large $|c|$ behavior
of Whittaker function, $M_{\h,\pm ic}(z)\sim z^{\h\pm ic},
M_{-\h,\pm ic}(z)\sim z^{\h\pm ic},$ when $|c|\rightarrow \infty$.
So $\lim_{|c|\rightarrow \infty}D(c,z)\sim 2z$. Using
$M_{\lambda,\mu}(z)=e^{-z/2}z^{\mu+\h}\Phi(\mu-\lambda+\h,2\mu+1;z)$,
and the relation
$\Phi(\alpha,\gamma;z)=e^z\Phi(\alpha-\gamma,\gamma;-z)$, we can
write $D(c,z)$ in terms of $\Phi$ as \eqn\aabs{D(c,z)=z
[\Phi(ic,2ic+1;z)\Phi(-ic,-2ic+1;-z)+\Phi(ic,2ic+1;-z)\Phi(-ic,-2ic+1;z)]
.}  $\Phi(\alpha,\gamma;z)$ as a function of $\gamma$ has single
poles at $\gamma=-n$, and analytic elsewhere. Near the pole,
$\lim_{2ic+1\rightarrow -n }\Phi(ic,2ic+1;z)\sim {(-1)^n\over
n!(2ic+1+n)}\left(\matrix{\h(n-1)\cr
n+1}\right)z^{n+1}\Phi({n+1\over 2},n+2;z)$, where
$$\left(\matrix{\h(n-1)\cr n+1} \right)\equiv
\h(n-1)[\h(n-1)-1]\cdots [\h(n-1)-n]/[(n+1)!].$$ If $n$ is odd, the
above is zero, so the potential poles in the upper half plane are at
$2ic+1=-2n$. However, \eqn\residue{\eqalign{&\lim_{2ic+1\rightarrow
-2n } D(c,z)\cr=&{z \over (2n)!(2ic+1+2n)}\left(\matrix{\h(2n-1)\cr
2n+1}\right) \Phi({2n+1\over 2},2n+2;z)\Phi({2n+1\over
2},2n+2;-z)\cr \times&
 [z^{2n+1}+(-z)^{2n+1}]=0.}}

The same phenomenon happens when $2ic+1\rightarrow 2n$. Thus
$D(c,z)$ is analytic in the complex plane as a function of $c$.
Since $D(c,z)$ approaches to $2z$ as the $|c|\rightarrow \infty$,
$D(c,z)=2z$ by Cauthy integral formula.

\listrefs
\end